\documentstyle[aps,preprint]{revtex}

\begin{document}

\textwidth=15.0cm
\textheight=23.0cm
\topmargin=-50pt
\oddsidemargin=0.5in

\title{Volatility clustering and scaling for financial time series due to attractor bubbling}
\author{A.\ Krawiecki$^{a,b}$ \and J.\ A.\ Ho\l yst$^{a,b}$ \and
and D.\ Helbing$^{b}$}

\address{
$^{a}$Faculty of
Physics, Warsaw University of Technology, Koszykowa 75, PL-00-662 Warsaw,
Poland}

\address{
$^{b}$Institute for Economics and Traffic, Dresden University of Technology,
D-01062 Dresden, Germany }

\baselineskip=4ex
\maketitle

\begin{abstract}
A microscopic model of financial markets is considered, consisting of many interacting
agents (spins) with global coupling and discrete-time thermal bath dynamics, similar to random
Ising systems. The interactions between agents change randomly in time.
In the thermodynamic limit the obtained time series of price returns show
chaotic bursts resulting from the emergence of attractor bubbling or on-off intermittency,
resembling the empirical financial time series with volatility clustering. For a proper choice
of the model parameters the probability distributions of returns exhibit power-law tails with
scaling exponents close to the empirical ones.
\end{abstract}

\vspace{0.3cm}


\newpage

In recent years, both the analysis and modeling of financial time series have attracted
a growing interest in statistical physics [1-11], motivated by a quest for a possibly
universal dynamics underlying different markets. Indeed,
various financial time series share similar properties, e.g., relative price changes (returns)
show intermittent occurrence of large bursts (volatility clustering), resulting in the 
power-law scaling of tails in their probability distributions, as well as of the 
autocorrelation function of their absolute values \cite{Gopikrishnan98}.
Large-scale microscopic simulations and analytic studies [4-11] revealed that such
scaling can result from phenomena as diverse as, e.g., percolation
\cite{Cont00} or stochastic processes with multiplicative noise \cite{Levy94,Takayasu97a,Takayasu97}.
It was also shown \cite{Lux99,Youssefmir97,Takayasu97a,Gaunersdorfer01} that in some models of market
dynamics volatility clustering resembles on-off intermittency,
an extreme kind of intermittency characterized by a sequence of large
chaotic bursts separated by almost quiescent laminar phases \cite{Platt93}, 
or attractor bubbling, the predecessor of on-off intermittency in the presence of additive
noise \cite{Ashwin94}. 

In this Letter we present simulations of financial time series based on a
microscopic model of many interacting agents with discrete-time dynamics. The
agents are treated as spins exposed to fluctuations (a "heat bath") and coupled
by randomly time-dependent Ising-like interactions. Related spin models, but with 
time-independent interactions, have been proposed for economic
\cite{Chowdhury99} and social systems, where they are theoretically and empirically
supported by the social impact theory of opinion formation \cite{Latane81}.
In contrast, we emphasize that the social interactions and the communication network among
individuals are time dependent. They are also more and more
determined by different kinds of long-range communication instead of spatial
neighborhood which makes the topological structure of the interaction network
unimportant. We show that, if the average strength of these interactions 
(which reflect the average reaction of agents to price changes) varies
randomly in time, the system displays volatility clustering, and probability
distributions of returns resemble those of empirical financial time series. A
peculiar feature of our model is that the dynamics on a microscopic level is
stochastic, but interactions between large numbers of agents cause the
macroscopic dynamics to become typical of dynamical systems with attractor
bubbling. In particular, in the mean-field approximation our model
reduces to a generic model for attractor bubbling. Another important feature is
the presence of various sources of stochasticity in the microscopic dynamics.
First, the agents make decisions (on investments) under the influence of 
the external environment and other agents. The variation of the strength of this
influence is simulated by explicit random updates of the interaction strengths
with the environment and among agents at every time step. Besides, the final
agent decision is determined according to a probabilistic rule that mimics the
the uncertainty in  human decision-making.

Explicitly, we consider $i=1,2,\ldots N$ agents (spins) with orientations
$\sigma_{i} (t) = \pm 1$, corresponding to the decision to sell ($-1$) or to buy
(+1) a share of a traded stock or commodity at
discrete time steps $t$. The orientation of agent $i$ at
time $t+1$ depends on the local field
\begin{equation}
\label{localfield}
I_{i} (t)= \frac{1}{N}\sum_{j=1}^{N} A_{ij}(t) \sigma_{j}(t) + h_{i} (t) \, ,
\end{equation}
where $A_{ij}(t)$ are time-dependent interaction strengths among agents
(reflecting their social or communication network), and $h_{i}(t)$ is an
external field reflecting the effect of environment (e.g., access to external
information which can differ between agents). The interaction strengths and
external fields change randomly in time,
\begin{eqnarray}
\label{strengths}
A_{ij}(t)&=& A\xi (t) + a \eta_{ij}(t) \, , \\
h_{i}(t) &=& h \zeta_{i} (t) \, , \nonumber
\end{eqnarray}
where, for simplicity, we assume that $\xi(t)$, $\eta_{ij}(t)$, and
$\zeta_{i}(t)$ are random variables with no correlation in space and time, 
uniformly distributed in the interval $( -1, 1)$. According to Eq.\
(\ref{strengths}), $A$ is a measure of the randomly varying average
interaction strength between agents, $a$ is a measure of the random deviations of 
the individual interaction strengths $A_{ij}(t)$ from the average, and $h$ is a measure
of the random influence of the environment.
The strengths $A_{ij}(t)$ can assume both positive and negative values,
corresponding to a tendency to imitate or avoid the orientation of the
interaction partners in the market. Below we show that the noises $\xi (t)$ and
$\zeta_{i} (t)$ are necessary to get results comparable to empirical
findings on price returns, while the effect of $\eta_{ij}(t)$ is averaged to
zero for large $N$. The orientations of all agents are updated
synchronously according to the probabilistic rule
\begin{equation}
\label{thermal}
\sigma_{i}(t+1)=
\left\{
\begin{array}{lll}
1 & \mbox{with probability} & p\\
-1 & \mbox{with probability} & 1-p,
\end{array}
\right.
\end{equation}
describing an {\em uncertainty in decision-making},
where $p =1/\left\{1+\exp\left[ -2I_{i}(t)\right] \right\}$ (analogously to the
well-known thermal bath dynamics). 

In the following we will be interested in modeling the time series and
distributions of the returns $G_{\Delta t} (t)
= \ln S(t) - \ln S(t-\Delta t)$, i.e., forward changes in the
logarithms of prices $S(t)$ over a time scale $\Delta t$.
Relative price changes are
proportional to the difference between demand and supply, i.e., between the number of buying
and selling decisions. Therefore, after
introducing the average orientation $x(t)= N^{-1} \sum_{i=1}^{N} \sigma_{i}(t)$
of agents we get $dS/dt \propto xS$, which after discretizing time yields
\begin{equation}
\label{returns}
G_{\Delta t} (t) \propto
\sum_{\tau=0}^{\Delta t-1} x\left( t-\tau \right) \, .
\end{equation}
In particular, $G_{1} (t) \propto x(t)$, which enables us to interpret the terms in Eqs.\ 
(\ref{localfield}) and  (\ref{strengths}) as follows: Since the average interaction strength
$A\xi (t)$ is common to all
connections, it is a measure of the average reaction of the agents to the price changes,
influencing their decisions via the term $A\xi (t) x(t)$ in the mean field $I_{i}(t)$. 
The terms $a\eta_{ij}(t)$ describe the {\em fluctuating interaction network} while
$h \zeta_{i} (t)$ describe the {\em fluctuating environment}. The local field in 
Eq.\ (\ref{localfield}) describes how agent $i$ anticipates price changes by averaging
the information on the opinions of his interaction partners and the external information
accessible to him. 

Before discussing the results of numerical simulations for large $N$, we study the mean-field
approximation of our model. For this purpose let us consider 
the case $A\neq 0$, $a=0$ and $\zeta_{i}(t) =\zeta (t)$ for all $i$. Then the
dynamics of $x(t)$ reduces to a one-dimensional map,
\begin{eqnarray}
x(t+1) &=& \tanh \left[ A\xi(t) x(t) + h \zeta(t)\right] 
\label{map}\\
& \approx & A\xi(t) x(t)+ h\zeta(t) \, ,
\nonumber
\end{eqnarray}
where the approximate equality holds for $\left| A\xi(t) x(t) +\right.$ $\left.
h\zeta(t)\right| \ll 1$. The map (\ref{map}) is a generic model  for on-off intermittency
if $h=0$ \cite{Platt93}, and for attractor bubbling if $h>0$\cite{Ashwin94};
let us also note that the approximate linearized form of Eq.\ (\ref{map})
belongs to a more general class of stochastic processes with multiplicative noise
\cite{Takayasu97a,Takayasu97,Kuramoto97}. In
the case $h=0$ and with $\xi(t)$ distributed uniformly in the interval $(-1,1)$,
the fixed point $x=0$ of Eq.\ (\ref{map}) loses stability for $A\ge \mbox{e} =
2.718\ldots$, as a result of the so-called blowout bifurcation, while for $h>0$
the stability is lost already for $A\ge 1$ \cite{Platt93,Ashwin94}. Above the
instability threshold the trajectory departs, on average, from zero, but,
due to the confining nonlinearity of the map (\ref{map}), 
after the departure is randomly reinjected towards $x=0$. Thus the
time series of $x(t)$ have intermittent character and consist of laminar phases,
during which $x(t)\propto {\cal O}(h)$, and chaotic bursts, during which
$x(t)\propto {\cal O} (1)$. Although the time series for on-off intermitency and
attractor bubbling look similar, there is a  qualitative difference between these phenomena:
In the case of on-off intermittency the bursts are caused solely
by the multiplicative noise $\xi(t)$ whose amplitude $A$ exceeds the blowout
bifurcation threshold, while in the case of attractor bubbling 
the multiplicative noise only amplifies to macroscopic sizes
small deviations from the fixed point caused by the additive noise $\zeta (t)$.
In Fig.\ 1(a), an example of time series $x(t)$ is
shown in the attractor bubbling regime. It bears certain resemblance to
empirical financial time series of price returns: There are frequent large
bursts, corresponding to volatility clustering, separated by long-lasting
laminar phases during which the dynamics of the map is governed mainly by the
additive noise $h\zeta(t)$. 

Another property of the map (\ref{map}) is that the probability distribution $\rho(x)$ of
$x$ (proportional to the probability distribution of $G_{1}$) shows pronounced
tails obeying power-law scaling. It was shown analytically
that for $h=0$ the distribution of $x$ following the linearized multiplicative 
stochastic process in discrete time (\ref{map}) obeys the power law scaling
$\rho(x) \propto x^{-\alpha-1}$ \cite{Kuramoto97}; 
this scaling remains valid also for $x\gg h>0$ at least
for Gaussian noises $\xi$, $\eta$ and continuous time \cite{Kuramoto97,Venkataramani95}. 
Following Ref.\ \cite{Kuramoto97} the exponent $\alpha$ can be obtained as a non-zero
solution of equation $\int_{-\infty}^{\infty} \exp\left(\alpha y\right) w(y) dy=
A^{\alpha} \left(1+\alpha\right)^{-1}=1$, where $y=\ln \left| A\xi\right|$ is a random
variable with density $w(y)= A^{-1}e^{y}$ in the range $y\le \ln A$.
This means that the cumulative distribution $P(x)$ exhibits power-law
tails with $P(x)\propto x^{-\alpha}$. The power law scaling of $\rho(x)$ and the 
analytic values of $\alpha$ (independent of $h$) are confirmed by our
numerical simulations (Fig.\ 1(b)). In order to obtain
a value $\alpha >2 $ (i.e., within the range observed in empirical stock market time
series \cite{Gopikrishnan98}), the parameter $A$ in Eq.\ (\ref{map}) should be
in the attractor bubbling regime $1< \left| A\right| <\mbox{e}$, so additive
noise with $h\neq 0$ is necessary. Moreover, the cumulative
distributions of normalized returns $G_{\Delta t}$ show tails obeying power-law
scaling with robust exponent $\alpha$ for time scales $\Delta t$ varying over more than one
order of magnitude (Fig.\ 1(c)), a fact also typical of financial time series
\cite{Gopikrishnan98}.

It is also known that, close to the blowout bifurcation, the power spectrum of
$\left| x(t)\right|$ from Eq.\ (\ref{map}) shows power law scaling 
$P(\omega) \propto \omega^{-1/2}$ for small $\omega$, suggesting a similar power law
scaling of the autocorrelation function of the absolute values of returns. However, in
the interesting attractor bubbling regime we observed only slow exponential decay of
the autocorrelation of $\left| x(t)\right|$.

Now, we turn to numerical multi-agent simulations of the system. Examples of
time series of the mean orientation $x(t)$ of agents are shown in Fig.\ 2(a). 
It can be seen that, as the number of agents increases, laminar
phases and chaotic bursts become more distinct, in contrast with
many other multi-agent models of stock market \cite{Egenter99}. In our
model, the laminar phases with $x\approx 0$ correspond to the disordered 
phase with no preferred orientation, and the bursts with $x
\propto {\cal O}(1)$ correspond to the ordered phase with a majority of agents
sharing one orientation ("herding behavior"). For large $N$, and even for $a>A$,
the time series of $x(t)$ resulting from the simulation of Eqs.\ (1) to (3) and of the 
mean-field Eq.\ (\ref{map}) look similar. Also the scaling exponents $\alpha$ are
close to those obtained from the mean field dynamics (\ref{map}) with the same
value of $A$. Note
that, in a system with a large but finite number of agents, there are always
fluctuations of $x(t)$ around zero in the disordered phase due to the heat-bath
dynamics of individual agents. These fluctuations
play the role of additive noise in Eq.\ (\ref{map}), enabling behavior
typical of attractor bubbling even if $h=0$ in Eq.\
(\ref{strengths}). 

The main outcome of our many-body simulation is that, in a system of many agents,
each of which is subject to "thermal bath" fluctuations, attractor bubbling can
appear as a collective phenomenon, as a result of time-dependent interactions
among agents. The requirement for its appearance is the average interaction
strength between agents to vary randomly in time around the value at which the
phase transition between the disordered and ordered state occurs. In the
thermodynamic limit, the macroscopic system dynamics reduces then to the
mean-field dynamics governed by the average interaction strength. The details of
the interactions among pairs of agents, and the details of the variation of the mean 
interaction strength \cite{Platt93}, seem unimportant for the qualitative result.

To summarize, in this Letter we generalized a model of social impact to the
case of randomly-in-time varying interactions among agents, and we have applied it to
simulations of financial time series. Certain empirical properties of these time 
series, like volatility clustering and a power law distribution of returns,
were shown to result from attractor bubbling, which
appears in our model in the limit of many interacting agents. Thus we perceive
the market as a dynamical system with an extreme kind of intermittency, in 
contrast with other models with time-dependent interactions between agents \cite{Cont00},
which are related rather to self-organized criticality. Due to the lack of 
topological structure of the interaction network, the mean field approximation of 
our model quantitatively reflects the multi-agent dynamics. The necessary
conditions for the applicability of the model are (i) random fluctuations in time of
the average interaction strength between agents (corresponding to their average
reaction to price changes), (ii) an uncertainty of decision making analogous to
thermal heat bath dynamics,
and (iii), in the thermodynamic limit, small additive noise simulating the effect of
the external environment.

The authors would like to thank the Quandt Foundation of the ALTANA AG for
financial support of the project
``Nonlinear Dynamics in Models of Complex Systems''.
This work was supported in part
by the Polsih Committee for Scientific Research under grant No.\ 5 P03B 007 21.

{\small
Fig.\ 1 (a) Time series of returns $x(t)$ from the map (\ref{map}) with $A=2$, $h=10^{-2}$;
(b) Probability distributions 
$\rho(x)\propto \rho(G_{1})$ of returns from the map (\ref{map}) and
fits to the power law scaling $\rho (x)\propto x^{-\alpha-1}$ of their tails for $h=10^{-3}$ and $A=1.6$
($\circ$, $\alpha= 2.67$, $\alpha_{an}=2.89$); 
$h=10^{-2}$ and $A=1.6$ ($\Box$, $\alpha=2.62$, $\alpha_{an}=2.89$), 
$A=2.0$ ($\triangle$, $\alpha=0.98$, $\alpha_{an}=1.0$), 
$A=2.75$ ($\Diamond$, $\alpha=-0.10$, $\alpha_{an}=-0.032$); $\alpha_{an}$ denotes analytic values of
$\alpha$.
(c) Cumulative distributions of normalized
returns $G_{\Delta t}$ (where the angular brackets denote time average) from the map
(\ref{map}) with $A=1.6$, $h=10^{-2}$, $\Delta t=1$ ($\circ$), $\Delta t=10$
($\Box$), $\Delta t=100$ ($\triangle$), $\Delta t=1000$ ($\Diamond$), and comparision
with the power law $P\left( G_{\Delta t}\right) \propto G_{\Delta t}^{-\alpha}$ with 
exponent $\alpha=2.76$.
All distributions were obtained from positive returns.

Fig.\ 2 (a) Time series of returns $x(t)$ from the model with many interacting agents with
$A=2.0$, $a=4.0$ and, from above, $N=25$ and $N=4000$; (b)
Probability distributions $\rho(x)$ of returns and
fits to the power law scaling $\rho (x)\propto x^{-\alpha-1}$ of their tails
from the model with $N=1000$, $h=0$, $a=2A$, and $A=1.6$ ($\Box$, $\alpha=2.59$), $A=2.0$
($\triangle$, $\alpha= 0.98$), $A=2.75$ ($\Diamond$, $\alpha=-0.13$).
All distributions were obtained from positive returns.
}

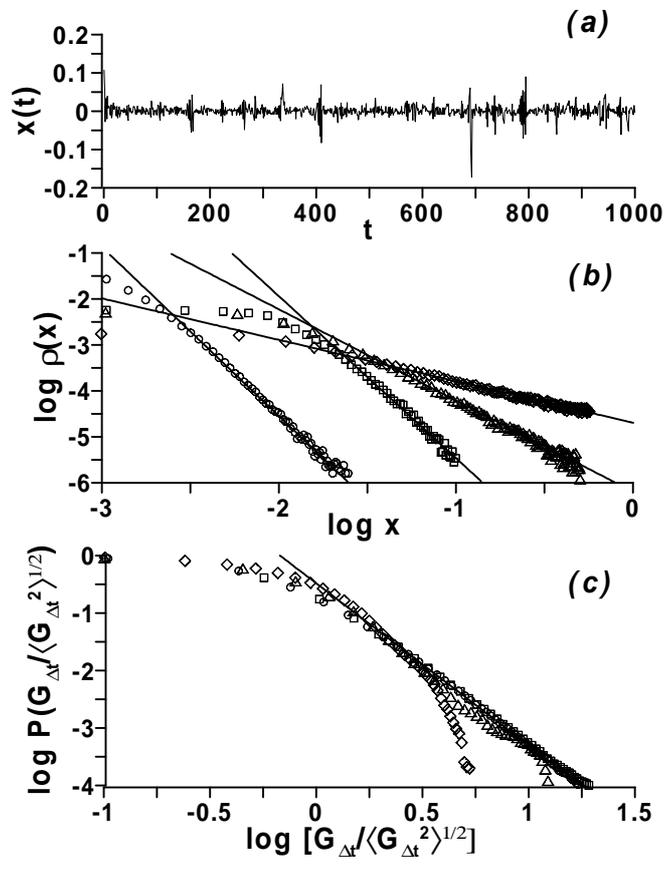
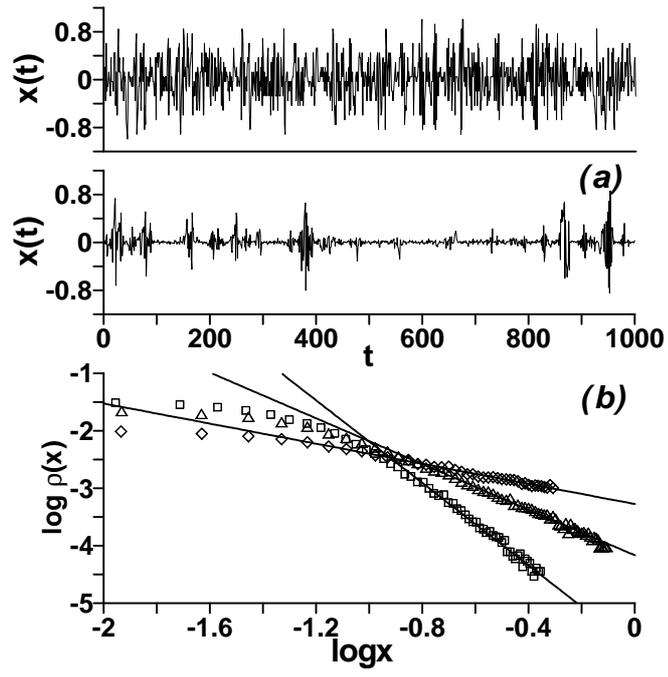
\end{document}